\RequirePackage{fix-cm}
\documentclass[smallextended]{svjour3}       
\smartqed  
\usepackage{graphicx}
\begin{document}

\title{Casting Light on Dark Matter\thanks{The work described here was supported partly by the London
Centre for Terauniverse Studies (LCTS), using funding from the European
Research Council via the Advanced Investigator Grant 267352.}
}


\author{John Ellis         
}


\institute{J. Ellis \at
              TH Division, Physics Department, CERN, CH-1211 Geneva 23, Switzerland, \&
Theoretical Particle Physics and Cosmology Group, Physics Department,
King's College London, London WC2R 2LS, UK\\
              \email{John.Ellis@cern.ch}           
}

\date{CERN-PH-TH/2011-141, KCL-PH-TH/2011-18, LCTS/2011-03}

\maketitle

\begin{abstract}
The prospects for detecting a candidate
supersymmetric dark matter particle at the LHC are reviewed, and compared
with the prospects for direct and indirect searches for astrophysical dark matter.
The discussion is based on a frequentist analysis of the preferred regions of the Minimal
supersymmetric extension of the Standard Model with universal soft
supersymmetry breaking (the CMSSM).
LHC searches may have good chances to observe supersymmetry in the near future
- and so may direct searches for astrophysical dark matter particles, whereas indirect
searches may require greater sensitivity, at least within the CMSSM.
\end{abstract}

\section{Introduction}
\label{sec:1}

The standard list of open questions beyond the Standard Model
of particle physics~\cite{COlex} includes the following.
(1) What is the origin of particle masses and, in particular, are they due to a Higgs boson?
(2) Why are there so many different flavours of standard matter particles, e.g., three neutrino species?
(3) What is the dark matter in the Universe?
(4) How can we unify the fundamental forces?
(5) Last but certainly not least, how may we construct a quantum theory of gravity?
Each of these questions will be addressed, in some way, by experiments at the LHC,
though answers to all of them are not guaranteed! I would argue that supersymmetry
is capable of casting important light on all but one of these questions, the exception being that
of flavour.
As you can guess from the title of this talk is, its focus is on question (3) concerning dark matter and,
in light of the above comments, specifically on ways to probe experimentally supersymmetric
models of dark matter. These include searches at the LHC as well as astrophysical dark
matter searches, and the presentation is based on my personal research in these areas: I
apologize to others for not mentioning adequately their work, which is referred to in the papers
referenced here.

\section{Supersymmetric Models}
\label{sec:2}

Supersymmetry is a very beautiful theory that is unique in its ability to link bosons and fermions and
hence, in principle, force and matter particles. However, phenomenological
interest in looking for supersymmetry was sparked by the realization that it could
stabilize the electroweak scale if supersymmetric partners or Standard Model
particles weigh $\sim 1$~TeV, and the flames of enthusiasm were fanned by the
subsequent realizations that in this case it could also facilitate unification of the 
fundamental interactions, would predict a light Higgs boson, and could explain the
apparent discrepancy between the experimental value of $g_\mu - 2$ and
theoretical calculations within the Standard Model~\cite{COlex}. Moreover, there
are general arguments that a dark matter particle that was once in equilibrium in
the early Universe, such as the lightest supersymmetric particle (LSP)~\cite{EHNOS} would
naturally have the right density to provide dark matter if it weighed $\sim 1$~TeV.

The LSP is stable in many supersymmetric models because of the multiplicative conservation of $R$-parity,
where $R = (-1)^{2S ÐL + 3B}$, $S$ denotes spin, and $L$ and $B$ are the lepton and 
baryon numbers. It is easy to check that Standard Model particles all have $R = +1$, 
whereas their supersymmetric partners would have $R = -1$. Multiplicative $R$ conservation would
imply that sparticles must be produced in pairs, heavier sparticles must decay into lighter sparticles,
and therefore the lightest supersymmetric particle (LSP) must be stable, and present in the
Universe today as a relic from the Big Bang. The LSP should have neither an electric charge nor
strong interactions, but should have only weak interactions. It is often thought to be the lightest
neutralino $\chi$, a mixture of the supersymmetric partners of the photon, $Z$ boson and neutral
Higgs bosons, and this identification will be assumed in the following. There are other possibilities
such as the gravitino, which would be difficult to detect astrophysically, but could also have distinctive
signatures at the LHC.

We work within the framework of the minimal supersymetric extension
of the Standard Model (MSSM), in which the known particles are
accompanied by their simple supersymmetric partners and there are two
Higgs doublets, with a superpotential coupling denoted by $\mu$ and
a ratio of Higgs v.e.v.s denoted by $\tan \beta$~\cite{MSSM}. The bugbear of the MSSM
is supersymmetry breaking, which may be parameterized generically through scalar
masses $m_0$, gaugino fermion masses $m_{1/2}$, trilinear soft
scalar couplings $A_0$ and bilinear soft scalar couplings $B_0$.
In our ignorance about them, the total number of parameters in the MSSM 
exceeds 100! For simplicity,
it is often assumed that these parameters are universal at the scale of
grand unification, so that there are common soft supersymmetry-breaking
parameters $m_0, m_{1/2}, A_0$, a scenario called the constrained
MSSM (CMSSM)~\cite{Kane}.

\begin{figure*}
\begin{center}
  \includegraphics[width=0.475\textwidth]{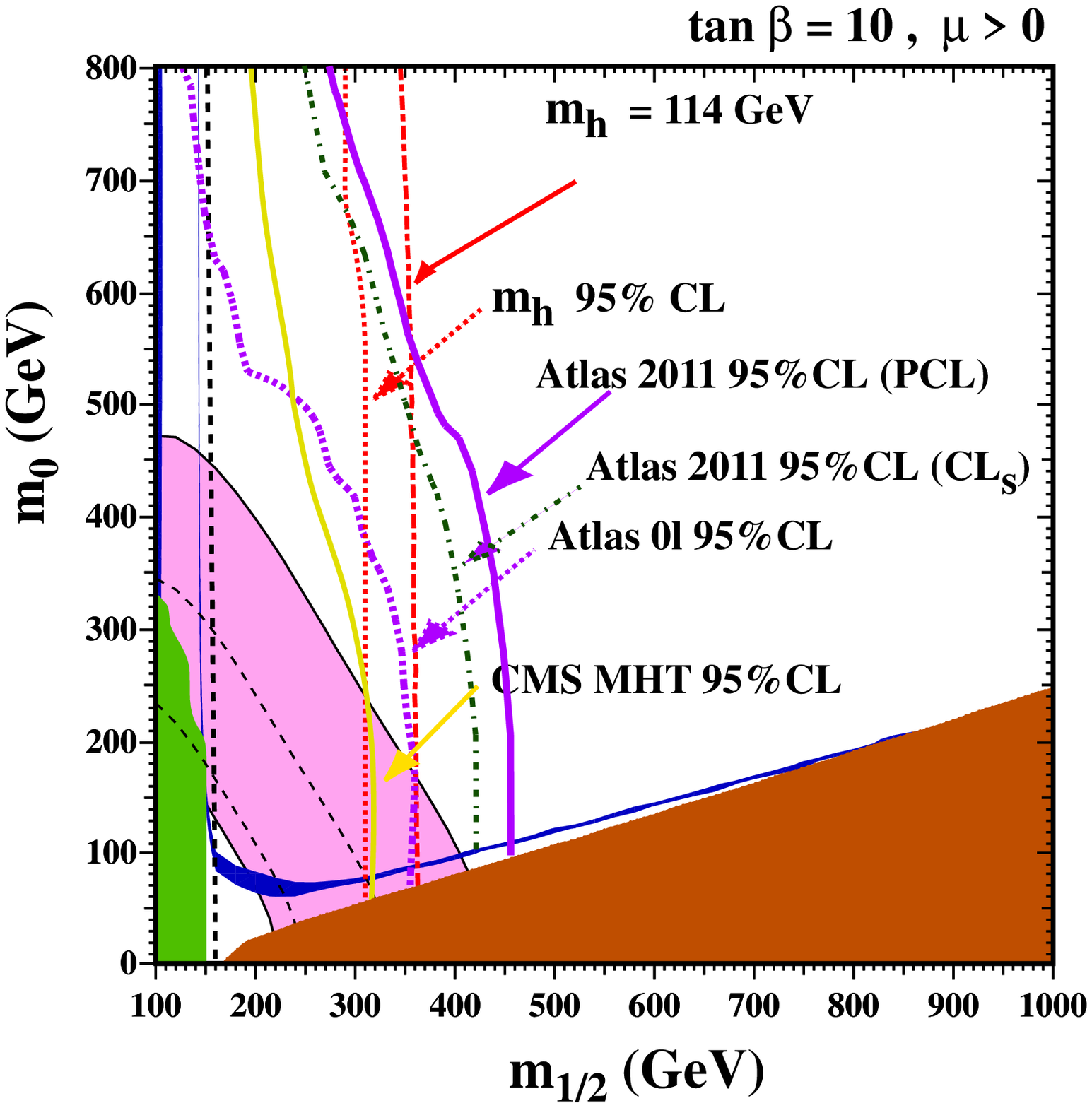}
    \includegraphics[width=0.475\textwidth]{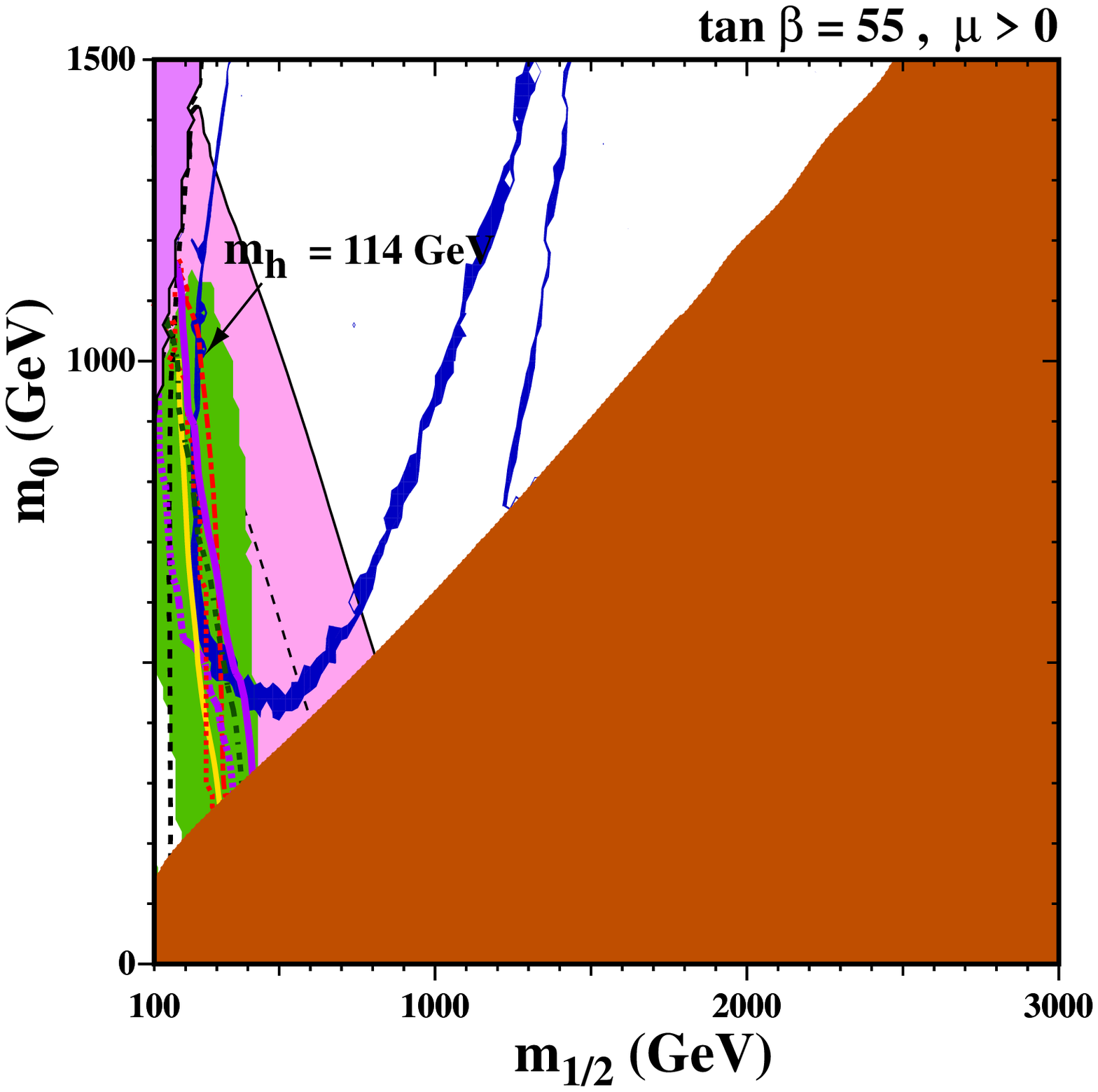}
\end{center}
\caption{The $(m_{1/2}, m_0)$ planes in the CMSSM for $\tan \beta = 10$ (left) and $\tan \beta=55$ (right),
assuming $\mu > 0$ and $A_0 = 0$~\protect\cite{EOSS,EOSgammas}, 
showing the 95\% CL limits imposed by ATLAS~\protect\cite{ATLAS,ATLAS165} and
CMS~\protect\cite{CMS} data (purple and yellow lines, respectively). The regions where the relic LSP
density falls within the range allowed by WMAP~\protect\cite{WMAP} and other cosmological observations
appear as strips shaded dark blue. The constraints due to the
absences of charginos~\protect\cite{LEPsusy} and the Higgs boson~\protect\cite{LEPH} at LEP are also shown, 
as black dashed and red dot-dashed lines,
respectively. Regions excluded by the requirements of electroweak symmetry breaking
and a neutral LSP are shaded mauve and brown, respectively. The green region is
excluded by $b \to s \gamma$, and the pink region is favoured by the supersymmetric
interpretation of the discrepancy between the Standard Model calculation and the
experimental measurement of $g_\mu - 2$~\protect\cite{E821} within $\pm 1$ and $\pm 2$ standard
deviations (dashed and solid lines, respectively).}
\label{fig:CMSSM}       
\end{figure*}

Fig.~\ref{fig:CMSSM} shows a compilation of theoretical, phenomenological, experimental
and cosmological constraints in two $(m_{1/2}, m_0)$ planes of the CMSSM, taking  as
examples $\tan \beta = 10, 55$ and $A_0 = 0$. (This plot updates the planes shown
previously in~\cite{EOSS,EOSgammas}. In addition to the absence of a stable charged
sparticle (which excludes regions at large $m_{1/2}$ and small $m_0$)
and the presence of a consistent electroweak vacuum
(which excludes regions at small $m_{1/2}$ and large $m_0$),
these constraints include the absences of supersymmetric particles at LEP 
(which require any charged sparticle to weigh more than
about 100~GeV~\cite{LEPsusy}), and at the Tevatron collider
(which has not found any squarks or gluinos lighter than about 400~GeV~\cite{Tevatron}). 
There are also important indirect
constraints implied by the LEP lower limit on the Higgs mass of 114.4~GeV~\cite{LEPH}, and the
agreement of the Standard Model prediction for $b \to s \gamma$ decay
with experimental measurements. The only possible strong experimental discrepancy
with a Standard Model prediction is for $g_\mu - 2$~\cite{E821}, though the significance
of this discrepancy is still uncertain. 
However, astrophysics provides a clear discrepancy with the Standard Model,
since dark matter cannot be explained without physics beyond the Standard
Model, such as supersymmetry. The fact that the dark matter density is
constrained to within a range of a few percent~\cite{WMAP}:
\begin{equation}
\Omega_{DM} \; = \; 0.111 \pm 0.006
\label{WMAP}
\end{equation}
constrains some combination of the parameters of any dark matter model also to within a few percent,
e.g., the parameters $(m_{1/2}, m_0)$ of the CMSSM for fixed values of its other parameters.

\begin{figure*}
\begin{center}
  \includegraphics[width=0.75\textwidth]{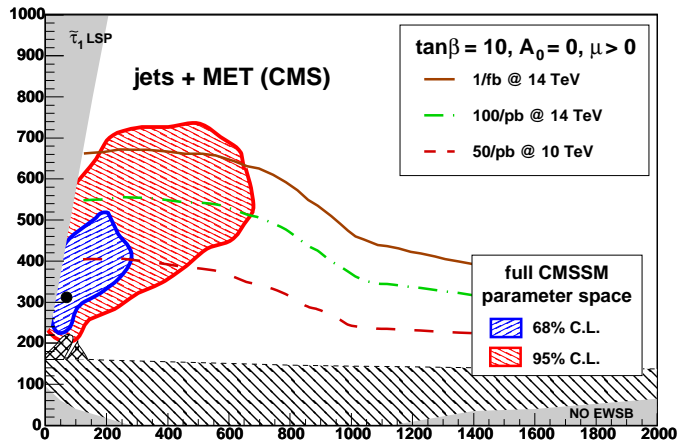}
\end{center}
\caption{Regions of the $(m_0, m_{1/2})$ plane favoured in a pre-LHC global frequentist analysis
of the CMSSM at the 68\% CL (blue) and the 95\% CL (red), as well as the best-fit point (black)~\protect\cite{MC2}.
Overlaid are the reaches estimated by CMS for discovering supersymmetry within the CMSSM,
with the indicated amounts of luminosity and centre-of-mass energy. These estimates were made
specifically for $\tan \beta = 10$ and $A_0$, but apply more generally.}
\label{fig:MC2}       
\end{figure*}

In a series of papers~\cite{MC2,MCn,MC6}, 
we have made global fits to the parameters of various supersymmetric models,
including those of the CMSSM, in a frequentist approach incorporating
precision electroweak data, the LEP Higgs mass limit, $\Omega_{DM}$,
data on the B decays $b \to s \gamma$, $B^\pm \to \tau^\pm \nu$ and $B_s \to mu^+ \mu^-$, and
(optionally) $g_\mu - 2$
One example of a pre-LHC frequentist fit within the CMSSM is shown in Fig.~\ref{fig:MC2}~\cite{MC2},
where the estimated reaches of the LHC experiments with various centre-of-mass energies and
integrated luminosities are indicated. These give cause for hope to discover supersymmetry in the
early days of the LHC, and we discuss later the implications of the first LHC searches. The
primary signals being sought at the LHC are events with missing transverse energy carried
off by dark matter particles, accompanied by hadronic jets and possibly leptons.

\section{Astrophysical searches for supersymmetric dark matter}

These include the direct search for elastic dark matter scattering on a nucleus in the laboratory,
$\chi + N \to \chi + N$, indirect searches for $\chi \chi$ annihilations in the core of the Sun or Earth
via energetic muons produced by energetic solar or terrestrial neutrinos, the indirect search for 
$\chi \chi$ annihilations in the galactic centre or elsewhere
via energetic $\gamma$ rays, and indirect searches
for $\chi \chi$ annihilations in the galactic halo via antiprotons or positrons among the cosmic rays.

\begin{figure*}
\begin{center}
  \includegraphics[width=0.75\textwidth]{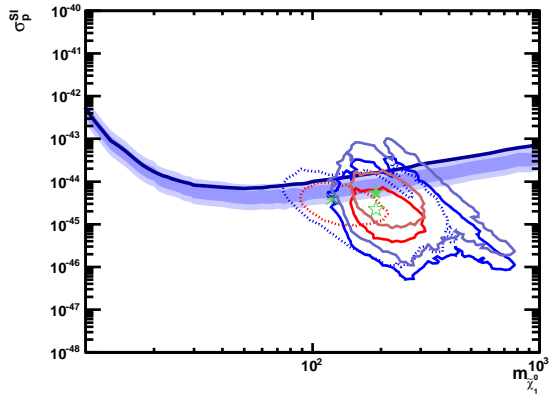}
\end{center}
\caption{The correlation found in the CMSSM between the spin-independent dark matter
scattering cross section $\sigma^{SI}_p$ 
and the LSP mass $m_{{\tilde \chi}^0_1}$ prior to the 2010 LHC data and the current Xenon100 results
shown by dotted lines, red and blue for  the 68 and 95\%~CL contours, respectively~\protect\cite{MC6}.
The solid lines include the 2010 LHC results: those calculated assuming $\Sigma_{\pi N} = 50$~MeV are
shown as brighter coloured curves and those for
$\Sigma_{\pi N} = 64$~MeV as duller coloured curves,
in each case disregarding uncertainties.
The green `snowflakes' (open stars) (filled
stars) are the best-fit points in the corresponding models. Also shown is the 90\% CL
Xenon100 upper limit~\protect\cite{Xenon100} and its expected sensitivity band.}
\label{fig:ssi}       
\end{figure*}

Pre-LHC predictions at the 68\% and 95\% CL for spin-independent
elastic dark matter scattering cross section on a proton, $\sigma^{SI}_p$, are shown
as dotted lines in Fig.~\ref{fig:ssi}~\cite{MC6}, and compared with the upper limit from the Xenon100
experiment~\cite{Xenon100} discussed below. We see that the best-fit prediction lies close to the current 
experimental sensitivity and within reach of prospective future experiments.

\begin{figure*}
\begin{center}
  \includegraphics[width=0.75\textwidth]{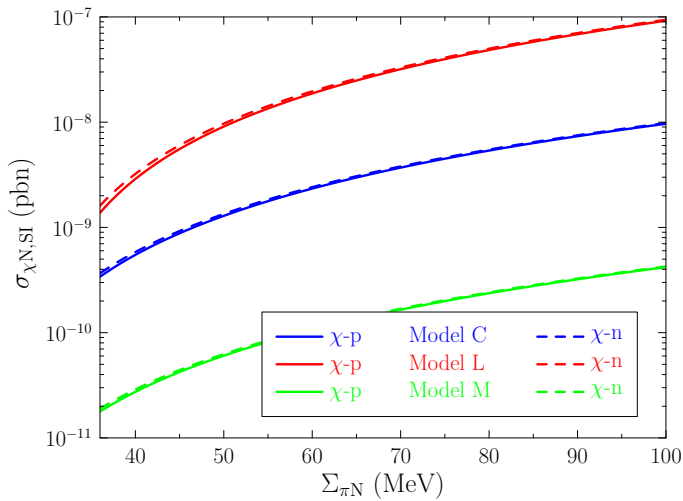}
\end{center}
\caption{Spin-independent dark matter scattering cross sections $\sigma^{SI}_{p,n}$ for three CMSSM
benchmark models C, L, and M~\protect\cite{EOS}. Note that $\sigma^{SI}_p$ and $\sigma^{SI}_n$ are 
nearly indistinguishable at the scale used in this plot.}
\label{fig:sigma}       
\end{figure*}

It should be noted, however, that these predictions are sensitive to the assumed value of the
$\pi$-N scattering $\sigma$ term, $\Sigma_{\pi N}$, as seen in Fig.~\ref{fig:sigma} for the cases
of some specific CMSSM benchmark scenarios~\cite{EOS}. This sensitivity arises because $\Sigma_{\pi N}$ is
sensitive to Higgs-exchange diagrams, which are sensitive, in turn, to the scalar density of strange
quarks in the proton: $\langle p | {\bar s} s | p \rangle$. Estimating this requires comparing
octet baryon mass differences, which yield a value for $\sigma_0 \equiv \frac{1}{2}(m_u + m_d) \langle p |
{\bar u} u +  {\bar d} d - 2 {\bar s} s | p \rangle = 36 \pm 7$~MeV, with $\Sigma_{\pi N} = \frac{1}{2}(m_u + m_d) 
\langle p | {\bar u} u + {\bar d} d \rangle$. The strangeness ratio $y \equiv  2\langle p | {\bar s} s | p \rangle/
\langle p | {\bar u} u  { \bar d} d \rangle = 1 - \sigma_0/\Sigma_{\pi N}$. Some experiments suggest a
relatively large value of $\Sigma_{\pi N} \sim 64$~MeV or more~\cite{Pavan}, and hence that $y$ is large,
whereas some lattice calculations suggest that $y$ is small~\cite{lattice}. Fig.~\ref{fig:ssi} displays as
brighter (duller) solid lines our post-2010-LHC predictions for $\sigma^{SI}_p$ assuming
$\Sigma_{\pi N} = 50 (64)$~MeV, demonstrating further the importance of
pinning down $\Sigma_{\pi N}$. In the following, we assume that
$\Sigma_{\pi N} = 50 \pm 14$~MeV, commenting on the implications if a larger value is assumed.

\section{Implications of initial LHC searches for supersymmetry}

\begin{figure*}
\begin{center}
  \includegraphics[width=0.75\textwidth]{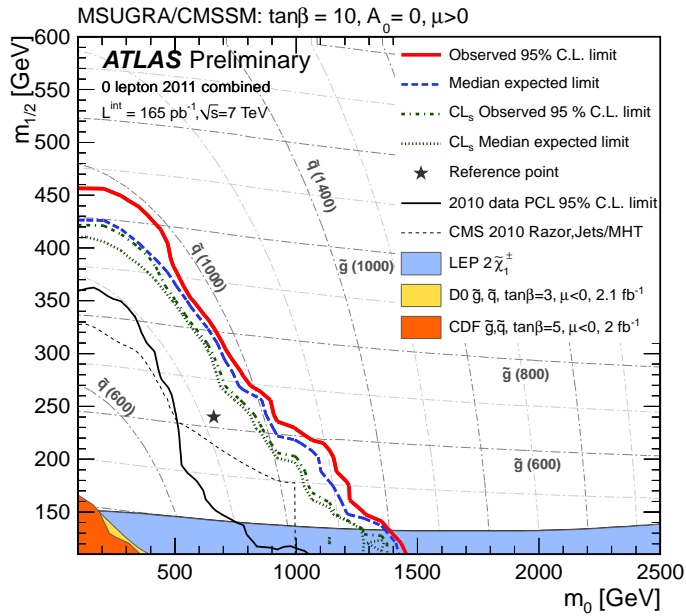}
\end{center}
\caption{Compilation of ATLAS and CMS 95\% CL exclusion limits in the $(m_0, m_{1/2})$ plane of the CMSSM. 
The ATLAS limit obtained with 165/pb of 2011 data, assuming $\tan \beta=10, A_0=0$ and $\mu > 0$ and
uisng the PCL (CL$_s$) method  is shown
as a solid red (dot-dashed green) line, also shown are the corresponding 
expected limits and a reference point~\protect\cite{ATLAS165}. 
The CMS~\cite{CMS} and ATLAS~\cite{ATLAS} limits from 2010 LHC data assuming $\tan \beta = 3$
are shown as dashed and solid black lines, respectively. Also shown for illustration are limits from the
Tevatron and from LEP.}
\label{fig:experiments}       
\end{figure*}

The CMS~\cite{CMS} and ATLAS~\cite{ATLAS} Collaborations have both announced negative results from initial
searches for supersymmetry using the $\sim 35$/pb of data each accumulated in 2010,
and ATLAS has also released preliminary results from a search using $\sim 165$/pb of 2011 data~\cite{ATLAS165}.
Their implications for the CMSSM may be represented in $(m_0, m_{1/2})$ planes
that are insensitive to $A_0$ and $\tan \beta$. As seen in Fig.~\ref{fig:experiments}, the
most sensitive constraints are those from searches for jets accompanied by missing
energy, which extend to much larger mass values than the previous Tevatron and LEP 
constraints on supersymmetry, as also seen in Fig.~\ref{fig:experiments}.

\begin{figure*}
\begin{center}
  \includegraphics[width=0.75\textwidth]{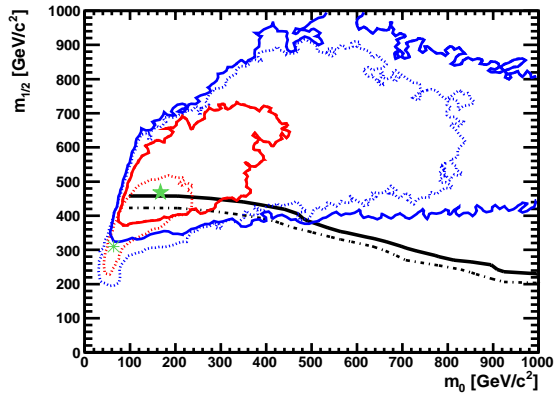}
\end{center}
\caption{The $(m_0, m_{1/2})$ plane in the CMSSM showing the regions favoured pre-LHC (dotted lines) and
after including all the 2010 LHC data as well as the Xenon100 limit (solid lines) at the 68\% CL (red) and
at the 95\% CL (blue)~\protect\cite{MC6}. Also shown in green are the pre-and post-2010-LHC/Xenon100 best-fit points,
and the preliminary 95\% CL limits obtained by ATLAS using 165/pb of 2011 data
using a PCL approach (solid black line) and a CL$_s$ approach (dash-dotted black line)~\protect\cite{ATLAS165}.}
\label{fig:LHCCMSSM}       
\end{figure*}

We have explored the implications of these data for supersymmetric models in an
extension of the frequentist global analysis introduced earlier~\cite{CMS}, including also the
negative results of LHC searches for the heavier supersymmetric Higgs bosons $H,A$~\cite{HA}
and also upper limits on $B_s \to \mu^+ \mu^-$ decay from the LHCb~\cite{LHCb}, CDF~\cite{CDF} and D0~\cite{D0}
experiments. Fig.~\ref{fig:LHCCMSSM} compares the best-fit points, 68\% and 95\% 
CL regions in the CMSSM before and after including the 2010 LHC data. We see that 
the best-fit point has moved to larger $m_{1/2}$, in particular, but still lies within the
pre-LHC 68\% CL region. Thus is not yet significant tension with the pre-LHC
predictions.

\section{The Xenon100 direct dark matter search experiment}

\begin{figure*}
\begin{center}
  \includegraphics[width=0.75\textwidth]{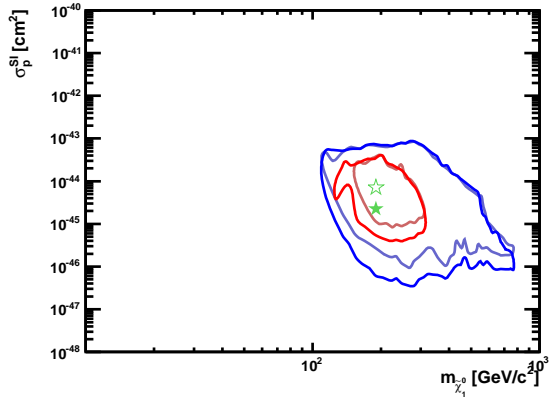}
\end{center}
\caption{The correlation found in the CMSSM between the spin-independent dark matter
scattering cross section $\sigma^{SI}_p$ 
and the LSP mass $m_{{\tilde \chi}^0_1}$ after including the constraints from
the 2010 LHC data and the Xenon100 results,
in red and blue for  the 68 and 95\%~CL contours, respectively~\protect\cite{MC6}.
Those calculated assuming $\Sigma_{\pi N} = 50 \pm 14$~MeV are
shown as brighter coloured curves and those for
$\Sigma_{\pi N} = 64 \pm 8$~MeV as duller coloured curves.
The green filled (open) star is the best-fit point for the corresponding range
of $\Sigma_{\pi N}$.}
\label{fig:ssi2}       
\end{figure*}

The Xenon100 experiment has recently announced results from an analysis of
100 days of data, establishing an upper limit on the spin-independent dark matter
scattering cross section $\sigma^{SI}_{p}$ that is significantly lower than those from previous experiments~\cite{Xenon100}.
As seen in Fig.~\ref{fig:ssi}, this
is the first dark matter scattering experiment that impinges significantly on the
expected parameter space of simple supersymmetric models such as the CMSSM.
However, the confrontation with CMSSM predictions must take into account the
uncertainties in the spin-independent hadronic scattering matrix element, primarily those
related to $\Sigma_{\pi N}$ that were discussed earlier. Fig.~\ref{fig:ssi2}
shows the results of including the Xenon100 results in the global CMSSM fit~\cite{MC6},
assuming the default option $\Sigma_{\pi N} = 50 \pm 14$~MeV (brighter colours). The main effect of
a larger value for $\Sigma_{\pi N}$ would be to increase the lower bound on
$\sigma^{SI}_{p}$, typically by a factor $\sim 3$ if $\Sigma_{\pi N} = 64 \pm 8$~MeV (duller colours). 

\section{Indirect Strategies for Detecting Supersymmetric Dark Matter}

\begin{figure*}
\begin{center}
  \includegraphics[width=0.75\textwidth]{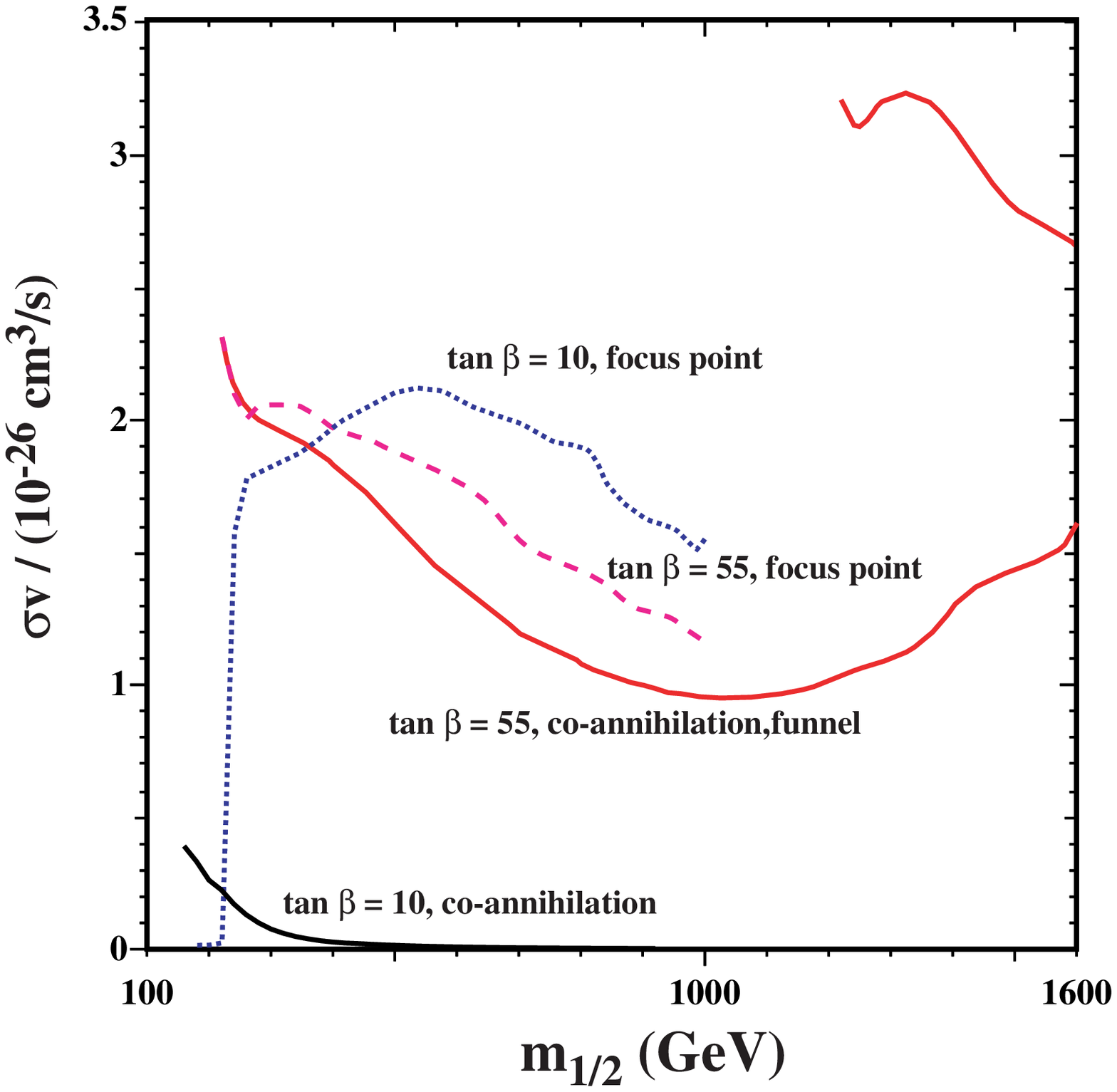}
\end{center}
\caption{The LSP-LSP annihilation cross section along the WMAP strips in the
coannihilation, focus-point and funnel regions for $\tan \beta = 10, 55$,
$A_0 = 0$ and $\mu > 0$, as functions of $m_{1/2}$. We see that the annihilation
cross section along the $\tan \beta = 10$ coannihilation strip is much smaller
than along the other strips, and decreases rapidly as $m_{1/2}$ increases~\protect\cite{EOSgammas}.}
\label{fig:annihilations}       
\end{figure*}

\begin{figure*}
\begin{center}
  \includegraphics[width=0.475\textwidth]{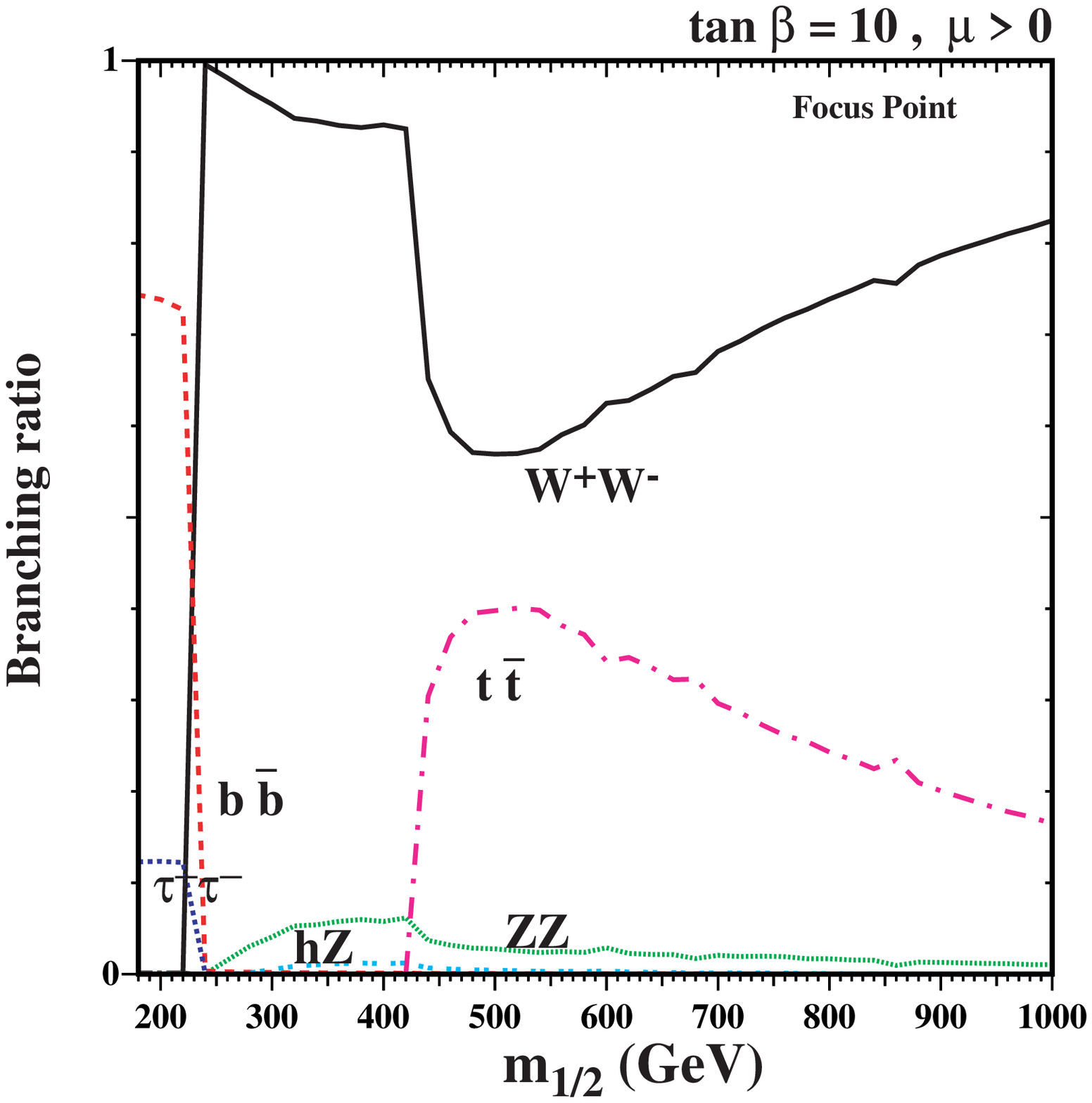}
  \includegraphics[width=0.475\textwidth]{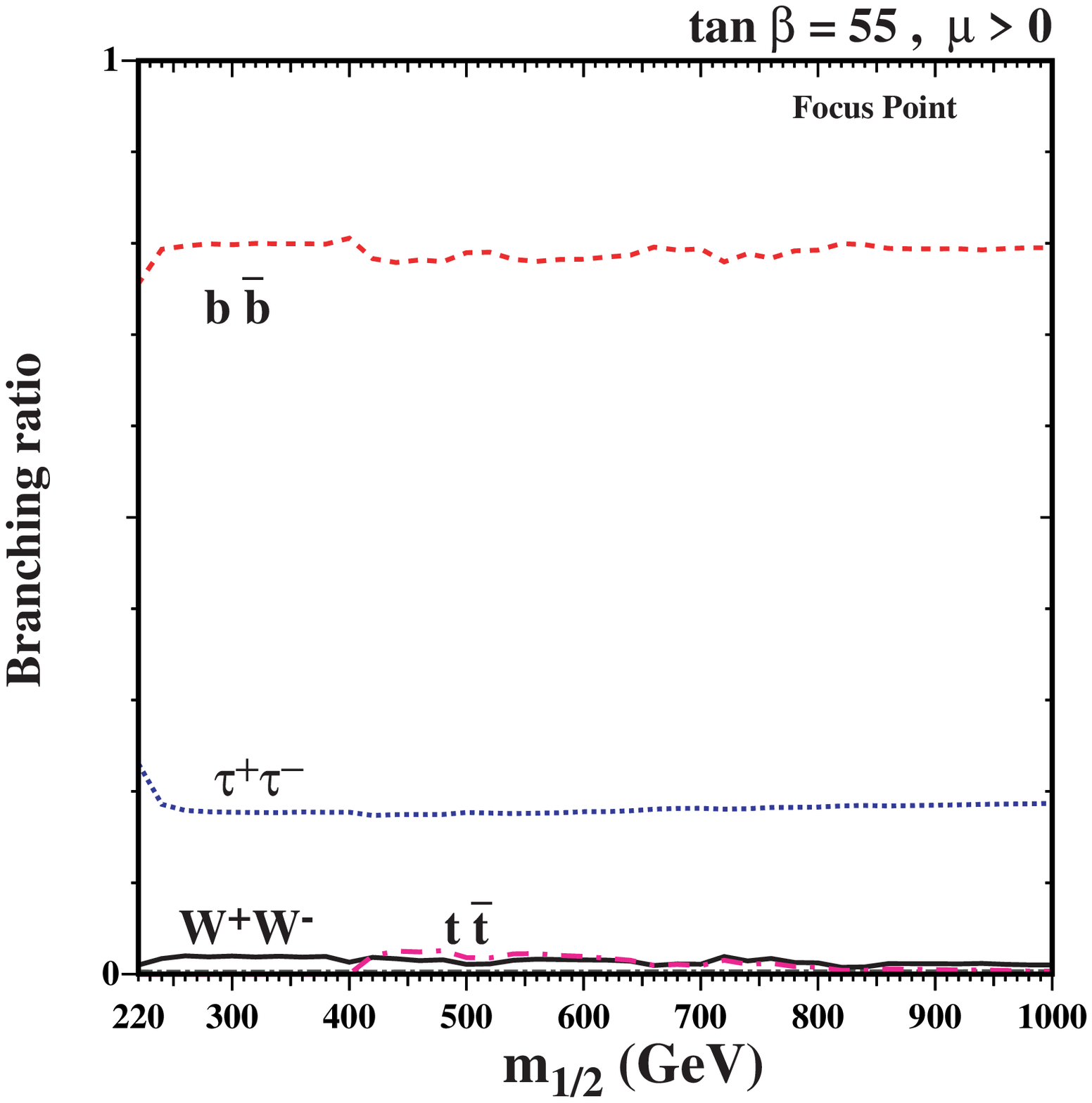}
\end{center}
\caption{The branching fractions for LSP-LSP annihilations into pairs of
Standard Model particles along the WMAP strips in the focus-point regions for $\tan \beta = 10$
(left) and $\tan \beta = 55$ (right)~\protect\cite{EOSgammas}.}
\label{fig:rates}       
\end{figure*}

These centre on searches for the products of dark matter annihilations at
low relative velocities, so the signals are all strongly dependent on the S-wave
annihilation cross section. Values along the WMAP strips for $\tan \beta = 10$
and 55 are shown in Fig.~\ref{fig:annihilations}~\cite{EOSgammas}. We see that the cross section
is generally much smaller along the coannihilation strip for $\tan \beta = 10$
than along the corresponding focus-point strip, or along both the WMAP strips
for $\tan \beta = 55$. Thus the latter strips offer {\it a priori} better prospects
the indirect detection of supersymmetric dark matter.
Discussions of indirect search strategies sometimes focus on annihilation final states
with particularly striking signatures such as $\chi \chi \to \gamma \gamma$. The
relative annihilation rates can be calculated in the CMSSM, as seen in Fig.~\ref{fig:rates}~\cite{EOSgammas}.
We see that the $\tau^+ \tau^-, W^+ W^-$ and ${\bar b} b$ final states dominate in
general, and that the $\gamma \gamma$ fraction is unfortunately very small in the CMSSM.

\subsection{Neutrino fluxes from dark matter annihilation in the Sun}

\begin{figure*}
\begin{center}
  \includegraphics[width=0.475\textwidth]{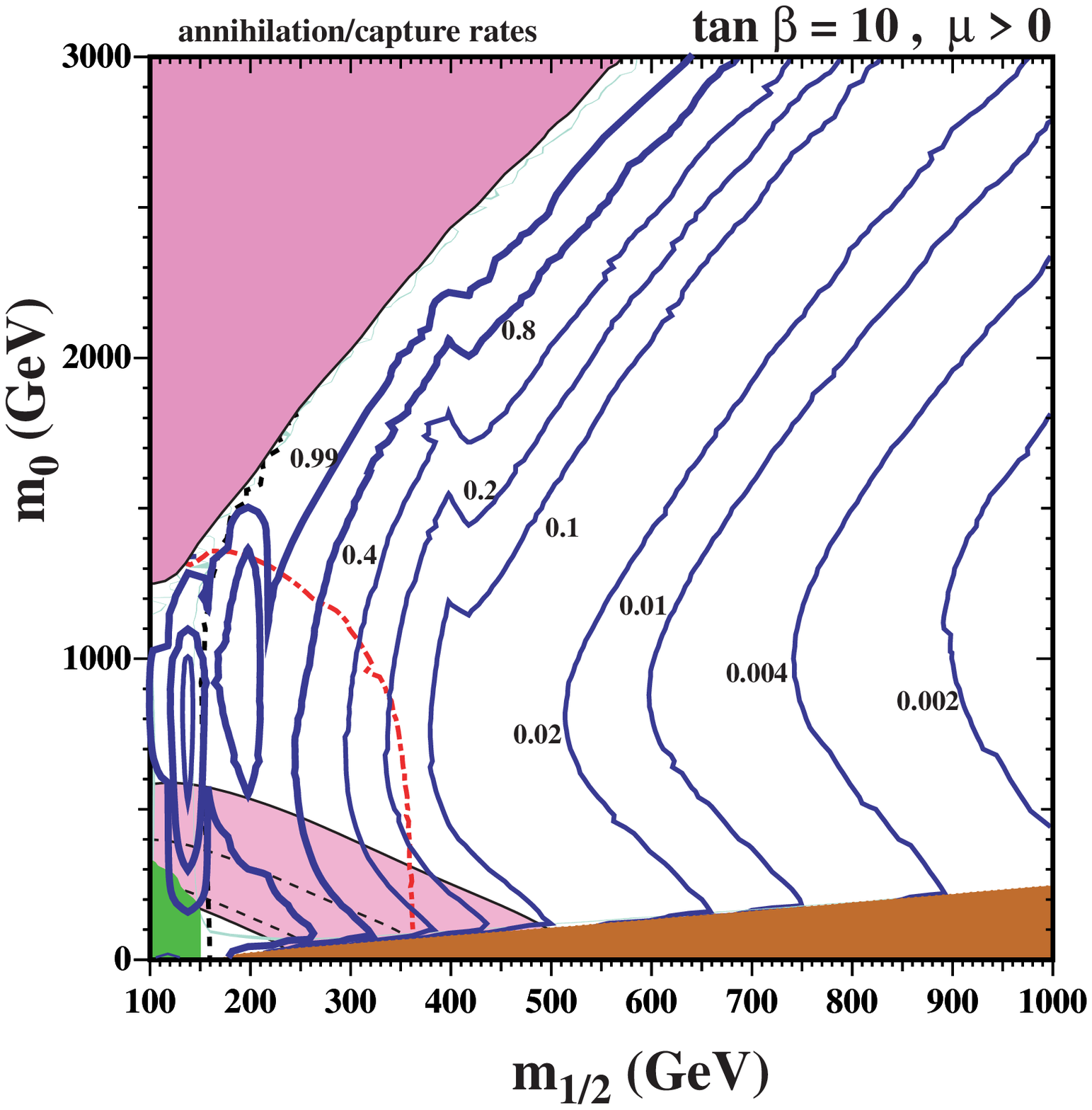}
  \includegraphics[width=0.475\textwidth]{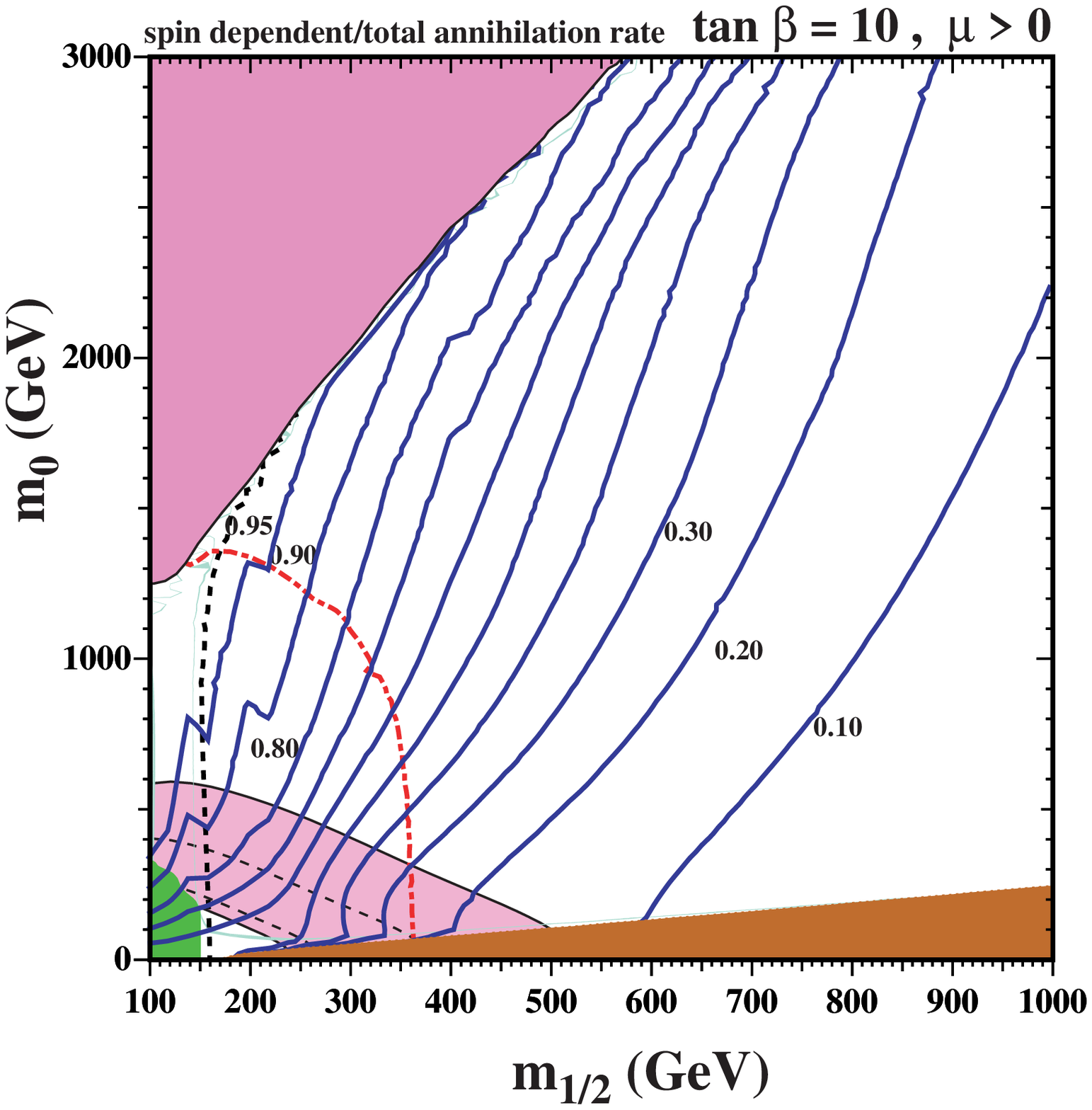}
\end{center}
\caption{The $(m_{1/2},m_0)$ plane in the CMSSM for $A_0 = 0$ and
    $\tan \beta = 10$, showing, in the left panel, contours of the
    ratio of solar dark matter annihilation and capture rates, as
    calculated using default values of the hadronic scattering matrix elements~\protect\cite{Savagenu}.
    Equilibrium corresponds to a ratio of unity, which is approached
    for small $m_{1/2}$ and large $m_0$. The right panel shows contours of the
    ratio of the solar dark matter annihilation rate calculated
    using only spin-dependent scattering to the total annihilation
    rate including also spin-independent scattering~\protect\cite{Savagenu}. It is seen that spin-independent
    scattering is important, even dominant, in much of the plane.}
\label{fig:Savage}       
\end{figure*}

Dark matter particles passing through the Sun or Earth may scatter, lose energy
and become gravitationally bound. These bound dark matter particles may then
annihilate, producing energetic neutrinos that generate detectable high-energy
neutrinos when they interact with matter in or near a detector.
It has often been assumed that the capture and annihilation processes are in
equilibrium, but this is not the case in the CMSSM in general, 
as seen for the $(m_{1/2}, m_0)$ plane with $\tan \beta = 10$ in the left panel of Fig.~\ref{fig:Savage}~\cite{Savagenu}.
It has also often been
thought that the dominant scattering mechanism in the Sun is spin-dependent scattering on
protons, but in studies of the CMSSM we have found that the dominant role may
actually be played by spin-independent scattering on heavier nuclei such as $^4$He, as seen in the right
panel of Fig.~\ref{fig:Savage}~\cite{Savagenu}.

\begin{figure*}
\begin{center}
  \includegraphics[width=0.95\textwidth]{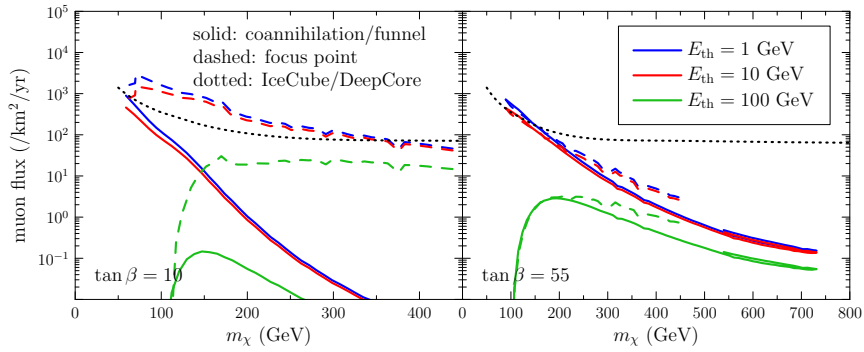}
\end{center}
\caption{The CMSSM muon fluxes though a detector calculated for $A_0 = 0$
    and (left) $\tan \beta = 10$, (right) $\tan \beta = 55$,
    along the WMAP strips in the coannihilation/funnel regions (solid)
    and the focus-point region (dashed)~\protect\cite{Savagenu}.
    Fluxes are shown for muon energy thresholds of (top to bottom)
    1~GeV, 10~GeV, and 100~GeV.
     Also shown is a conservative estimate of the sensitivity of the
    IceCube/DeepCore detector (dotted), normalized to a muon threshold
    of 1~GeV, for a particular hard annihilation spectrum
    ($\tau \bar{\tau}$ for $m_{\chi} < 80$~GeV, $W^+ W^-$ at higher
    masses).
    The IceCube/DeepCore sensitivity shown does not directly apply
    to the CMSSM flux curves, but can be treated as a rough
    approximation to which CMSSM models might be detectable.}
\label{fig:Savagestrip}       
\end{figure*}

Fig.~\ref{fig:Savagestrip} shows the resulting neutrino fluxes above various
neutrino energy thresholds along the WMAP strips in the CMSSM for $\tan \beta = 10$ and 55,
coampared with the possible sensitivity of the IceCube/DeepCore detector~\cite{Savagenu}.
The best prospects for detection are along the focus-point strip for $\tan \beta = 10$,
where the neutrino flux might be detectable in IceCube/DeepCore out to $m_\chi \sim 400$~GeV.
On the other hand, the prospects are quite unpromising along the corresponding coannihilation strip,
as one might have anticipated from the annihilation cross section plot in Fig.~\ref{fig:annihilations}. 
In the case of $\tan \beta = 55$,  there may be some prospects for neutrino detection for small
values of $m_\chi$ along the focus-point strip.

\subsection{Gamma-ray fluxes from dark matter annihilation in the Galactic core}

\begin{figure*}
\begin{center}
  \includegraphics[width=0.475\textwidth]{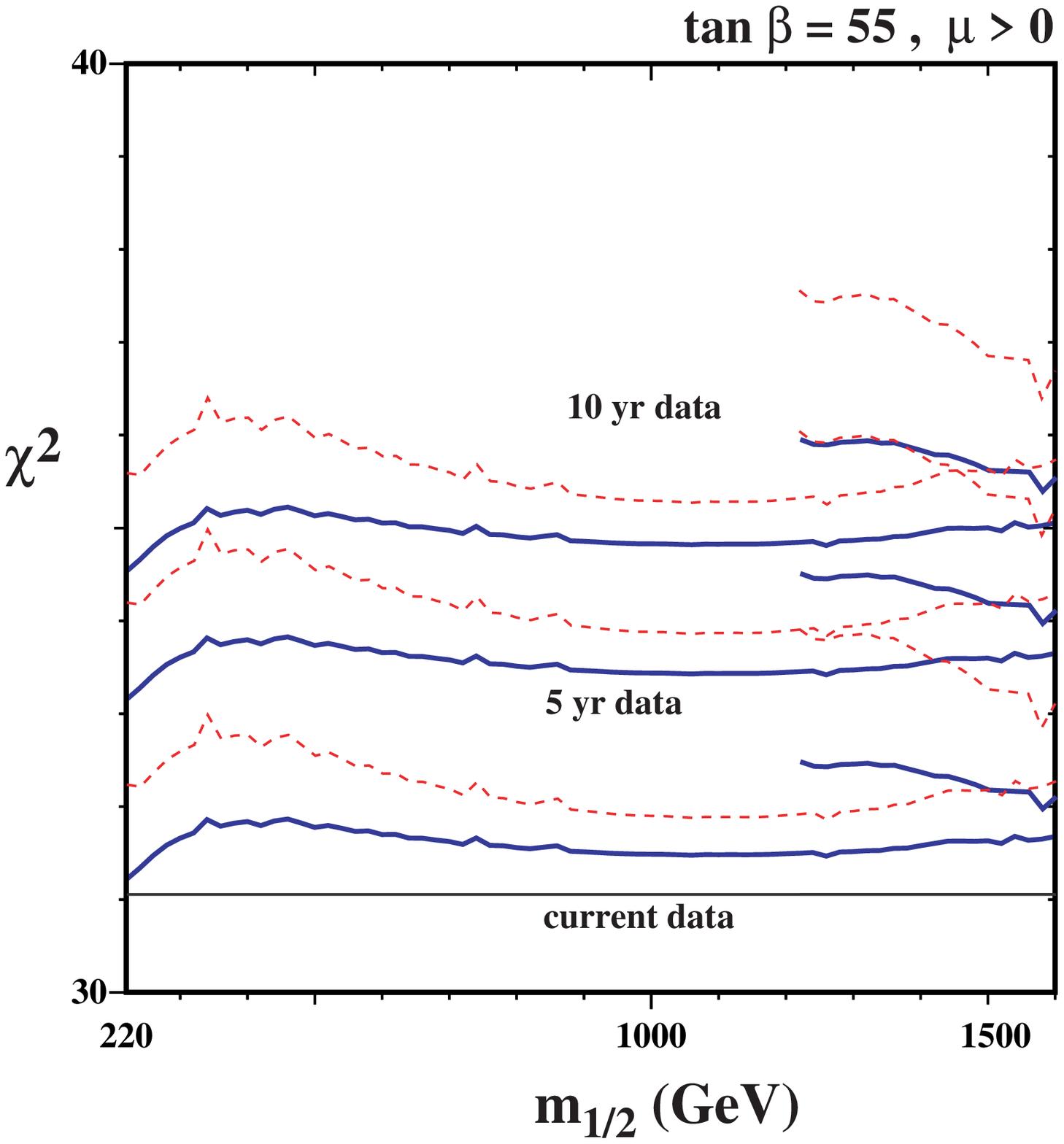}
  \includegraphics[width=0.475\textwidth]{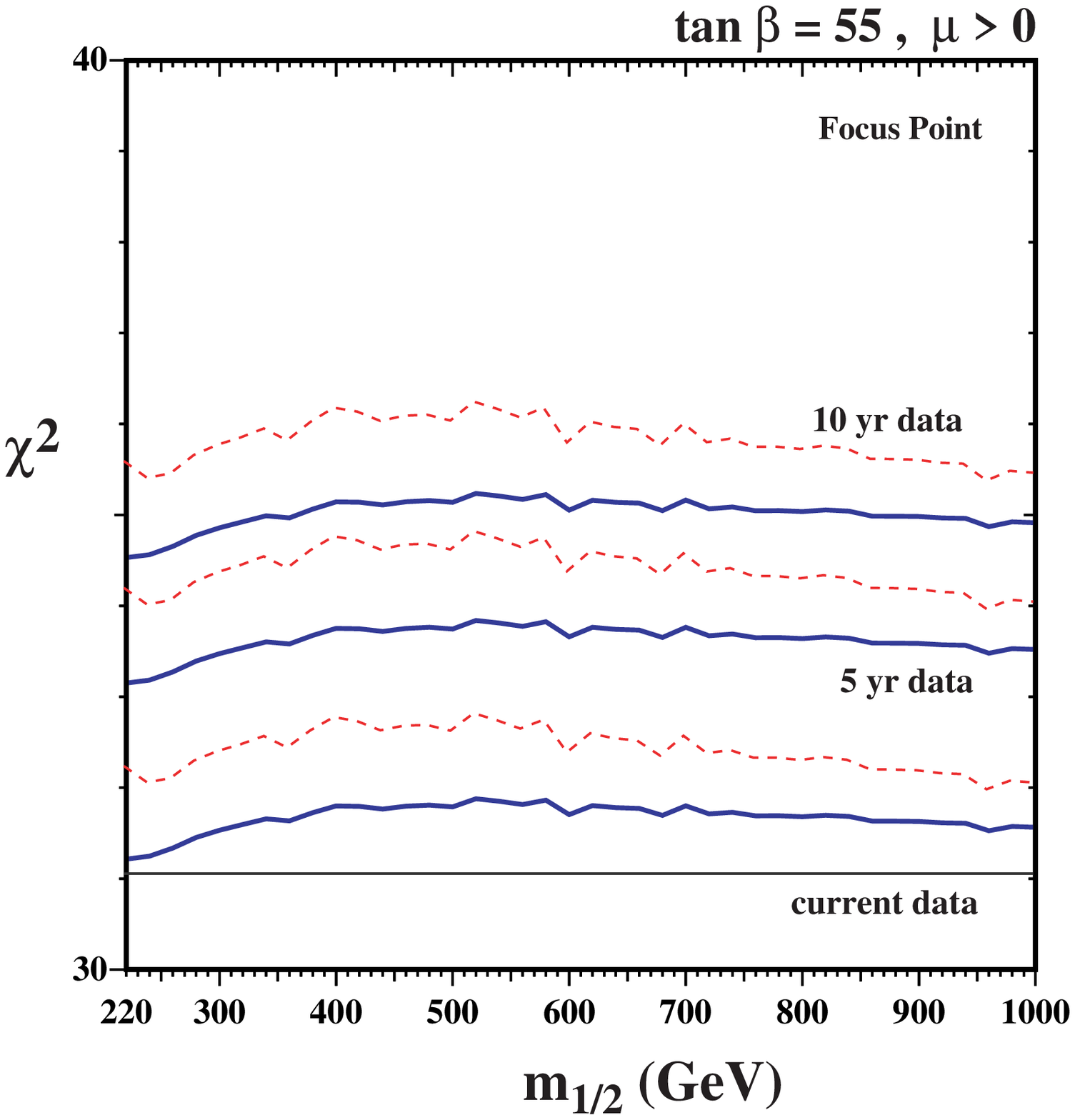}
\end{center}
\caption{The $\chi^2$ functions along the
CMSSM WMAP strips as functions of $m_{1/2}$  for $\tan \beta = 55$, along the 
coannihilation/funnel strip (left panel) and in the focus-point region
(right panel)~\protect\cite{EOSgammas}. In each panel, we display the $\chi^2$ function
for the background alone as a horizontal line at $\chi^2 = 31.1$,
the $\chi^2$ function obtained by adding the calculated LSP-LSP
annihilation signal in the current (approximately 2 1/2 year) Fermi data sample~\protect\cite{FLAT}, and
in projected 5- and 10-year data sets. Solid (blue) curves are based on an NFW profile~\protect\cite{NFW},
while dashed (red) curves are based on an Einasto profile~\protect\cite{Einasto}.}
\label{fig:gammas}       
\end{figure*}

Many studies have been made of the detectability of
high-energy $\gamma$ rays produced by the annihilations of dark matter particles
in the core of the Milky Way, in the Galactic bulge, in dwarf galaxies, and in the
diffuse cosmic background. Fig.~\ref{fig:gammas} displays the possible
sensitivity of the Fermi-LAT detector to $\gamma$ rays from dark matter
annihilations in the Galactic core along the coannihilation and focus-point strip
in the CMSSM with $\tan \beta = 55$~\cite{EOSgammas}. There is an important uncertainty associated
with the dark matter density profile in the core, and results are shown for a
Navarro-Frenk-White (NFW) profile~\cite{NFW} (solid blue lines) and for an Einasto~\cite{Einasto} profile
(dashed red lines). The lowest sets of lines in Fig.~\ref{fig:gammas} 
indicate the increase in $\chi^2$ that might arise
from a dark matter annihilation contribution, relative to a fit to the current Fermi-LAT data~\cite{FLAT} in
the absence of any supersymmetric signal, which corresponds to the horizontal solid black line 
with $\chi^2 = 31.1$. The higher sets of lines indicate what might be possible with 5- and 10-year
Fermi-LAT data sets: in these cases the values of $\chi^2$ in the absence of a supersymmetric signal
would be $\sim 33.0$ and 34.3, respectively.

\begin{figure*}
\begin{center}
  \includegraphics[width=0.475\textwidth]{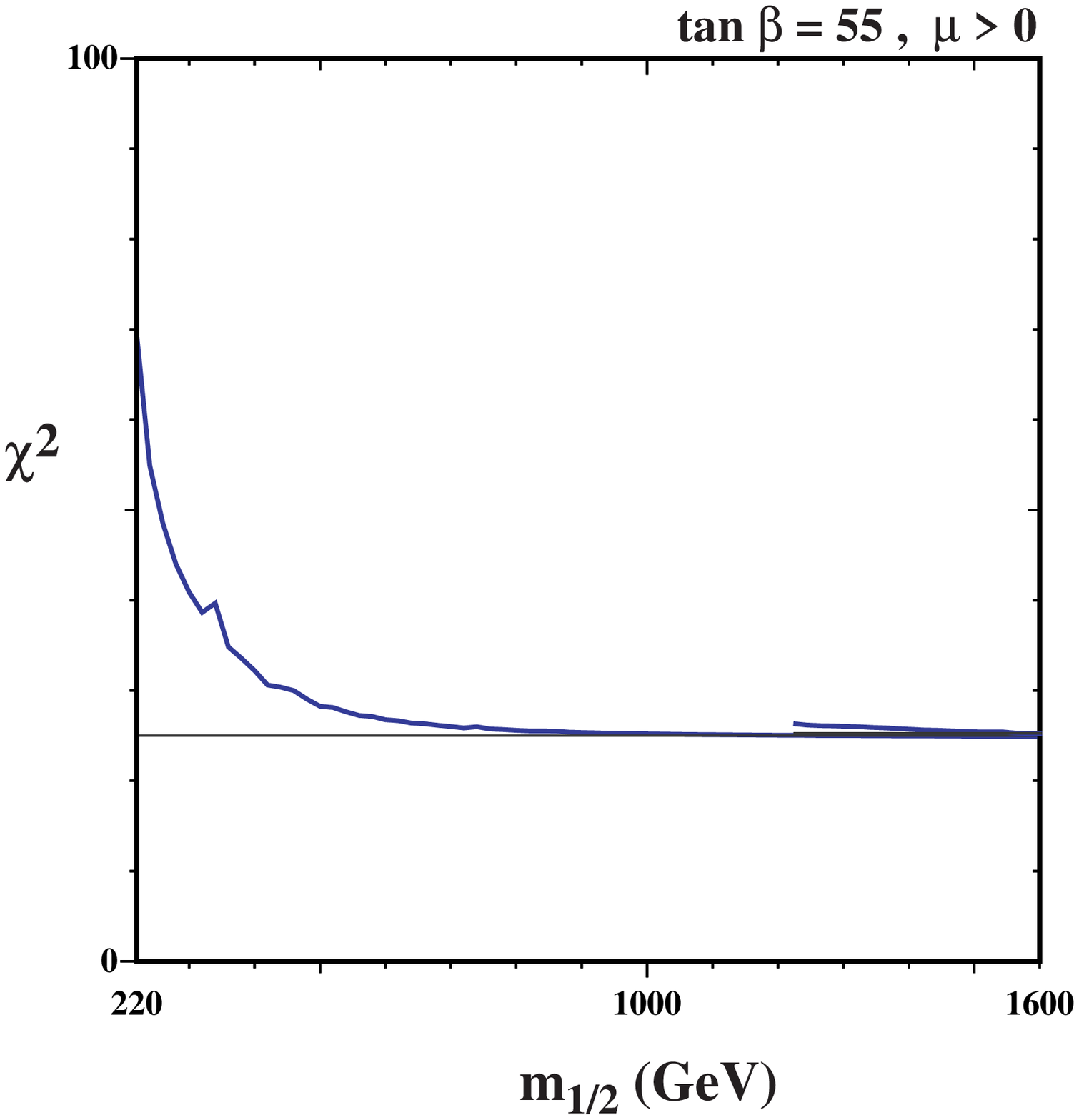}
  \includegraphics[width=0.475\textwidth]{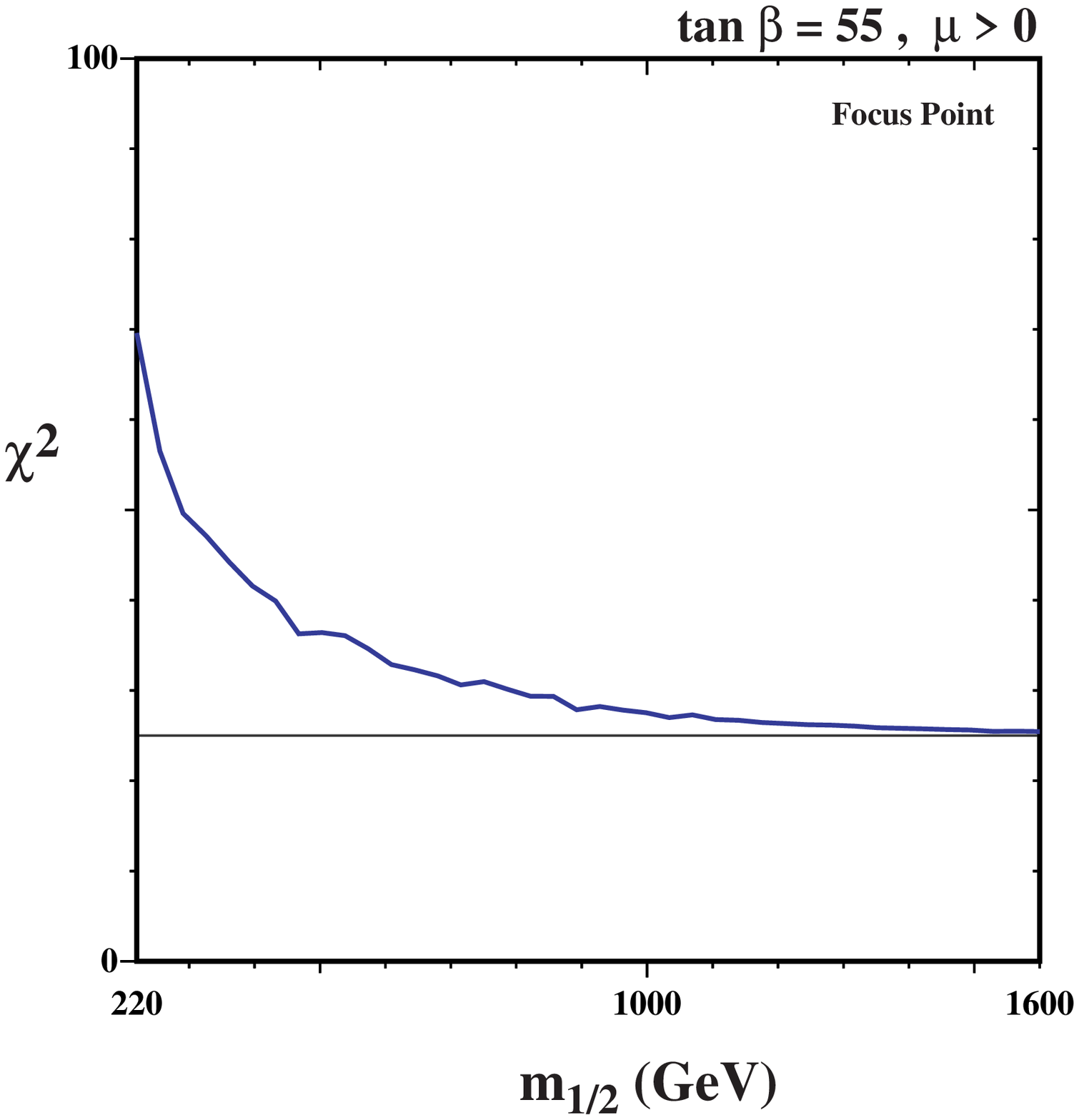}
\end{center}
\caption{The $\chi^2$ functions as functions of $m_{1/2}$ along the
CMSSM focus-point WMAP strip for $\tan \beta = 55$ assuming a negligible
systematic error, assuming that an improved
estimate of the background brings it to $\pm 1 \sigma$ from the data~\protect\cite{EOSgammas}. 
The left (right) panel is for the coannihilation/funnel strip (focus-point strip).}
\label{fig:gammas2}       
\end{figure*}

The sensitivity of the Fermi-LAT experiment could be increased significantly
with improvements in the understanding of the background and the systematic
uncertainty in the area of the detector. Fig.~\ref{fig:gammas2} indicates the
discrimination that could be possible if the background could be understood with
random errors of $\pm 1 \sigma$ relative to the measurement, and the
systematic error could be reduced to a negligible level. The blue lines assume
an NFW profile, and the sensitivity would be substantially
 increased if the core had an Einasto profile.
 
 \subsection{Anomalies in $e^\pm$ spectra?}
 
 A couple of surprising features have been observed in the $e^\pm$ spectra in the cosmic rays.
 One is a shoulder in the sum of the $e^+$ and $e^-$ spectra between $\sim 100$ and 1000~GeV~\cite{shoulder}, 
 and the other is an increase in the $e^+/e^-$ ratio between $\sim 10$ and 100~GeV~\cite{rise}. Before
 jumping to an interpretation involving the annihilations of dark matter particles, one should first
 consider more prosaic interpretations, taking into account uncertainties in cosmic-ray production and
propagation through the galaxy, as well as possible contributions from nearby sources. These may
render unnecessary an explanation in terms of dark matter annihilations, which would in any
case require a rather special supersymmetric model.

\subsection{Antiprotons and antideuterons from dark matter annihilations?}

The spectrum of cosmic-ray antiprotons has now been measured quite accurately,
and at energies $< 10$~GeV it seems to agree well with calculations of the production of secondary
antiprotons by primary matter cosmic rays~\cite{pbars}. However, there may be some discrepancy at 
higher energies, that could possibly be due to dark matter annihilations, though a more
conservative interpretation in terms of primary antimatter production by cosmic-ray sources
is also possible. Another possible signal of dark matter annihilations may be provided by
antideuterons with energies below about 1~GeV~\cite{antideuterons}, which may be distinguishable from
conventional antideuteron mechanisms that give a spectrum peaked at energies $> 1$~GeV.

\subsection{AMS}

The AMS experiment was launched successfully shortly after this talk, and subsequently
placed on the International Space Station~\cite{AMS}. We hope that it will be able cast light on the
dark mysteries mentioned in the two previous subsections!

\section{Final Remarks}

Hopefully this talk has convinced you that, on the one hand, the LHC may soon be
casting light on the nature of dark matter while, on the other hand, astrophysical
experiments may soon be casting light on fundamental questions in particle physics.
Only the synthesis between the two will have any chance of determining the true
nature of dark matter. Please also recall that important contributions to unravelling
this physics may be played by humble low-energy experiments, e.g., on $\pi - N$
scattering and pionic atoms, that may remove crucial uncertainties in this synthesis.
Please also note the important roles that may be played by antiparticles, including
antideuterons and positrons, as well as antiprotons. Finally, be ready for the
unexpected! We should all hope that the LHC will become famous for discovering
some unforeseen physics, and be open to the possibility that some unheralded
experiment will make the crucial discovery leading to a breakthrough in the
understanding of dark matter.




\end{document}